\documentclass[a4paper,11pt]{article}
\pdfoutput=1 

\usepackage{jcappub} 

\usepackage[T1]{fontenc} 

\title{\boldmath Measurement of Hubble constant: Non-Gaussian Errors in HST Key Project Data}

\author[a]{Meghendra Singh}
\author[b]{Shashikant Gupta}
\author[b]{Ashwini Pandey}
\author[c]{Satendra Sharma}

\affiliation[a]{Dr.A.P.J.Abdul Kalam Technical University, Uttar Pradesh, Lucknow 226021, India.}
\affiliation[b]{Amity University Haryana, Gurgaon, Haryana 122413, India.}
\affiliation[c]{Yobe State University, Damaturu, Yobe State, Nigeria.}

\emailAdd{meghendrasingh\_db@yahoo.co.in}
\emailAdd{shashikantgupta.astro@gmail.com}
\emailAdd{satyamkashwini@gmail.com}
\emailAdd{ssharma\_phy@yahoo.co.uk}

\abstract{Assuming the Central Limit Theorem, experimental uncertainties in any data set are expected 
to follow the Gaussian distribution with zero mean. We propose an elegant method based on 
Kolmogorov-Smirnov statistic to test the above; and apply it on the measurement of Hubble constant 
which determines the expansion rate of the Universe. The measurements were made using Hubble Space 
Telescope. Our analysis shows that the uncertainties in the above measurement are non-Gaussian. 
}

\begin{document}
\maketitle
\flushbottom

\section{Introduction}
Uncertainties are inevitable outcome of any experiment and their role is to spread the measured 
value around the true value of the quantity being measured. If the experiment is free of systematic 
effects, one expects the uncertainties to be symmetrically distributed around zero. Further, if Central 
Limit Theorem holds, they should follow Normal distribution.The systematics, if present, have to 
be identified and removed separately. Treatment of the errors becomes more important in astronomy 
since sometimes it is hard to repeat or perform the experiments in controlled way unlike the 
laboratory experiments. In the present letter 
we propose an elegant way to use the Kolmogorov-Smirnov (hereafter KS) test to detect the 
non-Gaussian uncertainties in astrophysical data, and apply it in the measurements of Hubble 
Constant.

In standard Big-Bang cosmology, the universe expands according to the Hubble law, $v=H_0d$, 
where $v$ is the recessional velocity of a galaxy at a distance $d$, and $H_0$ is the Hubble 
constant, which determines the expansion rate at the current epoch. Since, the velocities are 
measured in km/s and distances in Mega parsec (Mpc), the common unit of $H_0$ is km/s/Mpc and 
till the mid-1990s, most of the measured values fall in the range $40 \le H_0 \le 100$ km/s/Mpc 
\cite{tanvir95}. 

The value of Hubble constant is of fundamental importance for testing the framework of standard 
cosmology. It sets the age of the universe, size of the observable universe and defines the critical 
density of the universe, $\rho_c = 3H_0^2/8 \pi G$. Further, growth of structures in the universe also 
depend on the expansion rate, i.e., numerical value of $H_0$. The determination of many physical 
properties of galaxies and quasars (e.g., mass, luminosity, energy density) all require knowledge of 
the Hubble constant. Thus, determining the accurate value of $H_0$ is amongst the most important issues 
in cosmology.

More than eight decades have passed since Hubble (1929) initially published the Hubble law, however, 
pinning down the accurate value for the Hubble constant has been proved to be extremely challenging. 
The main difficulty lies in the measurement of accurate distances over cosmological scales. 

Hubble Space Telescope (HST) was launched in 1990 to measure the Hubble constant accurately. A space 
observatory was required since atmospheric seeing does not allow to resolve the Cepheids and measure 
their period-luminosity relations to large distances. The high resolution imaging of HST extends 
this limit, and the effective search volume. It has several other advantages as well, e.g., 
observations can be scheduled independently of the phase of the Moon, the time of day, or weather, and 
there are no seeing variations. 

One of the main key projects of HST was to measure the value of $H_0$ within 10\% accuracy, based on 
Cepheid calibration of a number of secondary distance determination methods. Determining $H_0$ 
accurately requires the measurement of distances far enough away so that both the small- and large-scale 
motions of galaxies become small compared to the overall Hubble expansion. To extend the distance scale 
beyond the range of the Cepheids, a number of methods that provide relative distances were chosen. The 
HST Cepheid distances were used to provide an absolute distance scale for these otherwise independent 
methods, including the Type Ia supernovae (SNe Ia), the Tully-Fisher relation (TF), the fundamental plane (FP) for elliptical 
galaxies, surface brightness fluctuations(SBF), and Type II supernovae (SNe II). The final result of HST key project was 
published in \cite{hstkey} (hereafter F01). However, some issues related to the HST key project data have 
also been reported. \citep{mccl07} found statistically significant spatial variation in the value of $H_0$, 
indicating the directional anisotropy. The variation does not appear to be an artifact of the Galactic dust; 
and the overall structure in the map is not consistent with the distribution of dust in the Cosmic Background 
Explorer (COBE) map \citep{schlegel98}. Using techniques based on extreme value theory \citep{gupta10}, 
\citep{gup11} have reported that the errors in HST key project data are non-Gaussian.

Our main task in this paper is to determine whether or not, the measurement errors in the HST key project 
data are Gaussian in nature?

\section{HST Key Project compilation}
It is natural for secondary distance indicators to be affected by their own systematic uncertainties. 
In order to use Cepheid calibration to a secondary method, one has to choose number of calibrating 
galaxies for a given method initially such that the final statistical uncertainty on the zero point 
for that method remain constrained to 5\%. Prior to HST, number of such calibrating galaxies were very 
small, e.g., only five for Tully-Fisher relation, none for SNe Ia, one for surface brightness fluctuations, 
and none for Fundamental plane relation.

For the calibration of secondary methods of Key Project, Cepheid distances of 18 new galaxies  were obtained, 
HST data for eight other galaxies were reanalyzed; and these distances were combined with the five other nearby
galaxies. Thus a total of 31 calibrating galaxies were available to serve the purpose as shown in Table 2 of 
F01. The maximum distance of calibrating galaxies for each secondary method in pre-HST \& post-HST era is shown in 
Table~\ref{tbl:F01}. It is clear from Table~\ref{tbl:F01} that the distance to the farthest calibrating galaxy prior 
to HST is 3.7 Mpc, while in the post-HST era it is more than 20 Mpc. These galaxies were observed in the active star forming regions of sky,  but low in apparent dust extinction. 
Observations carried in two different wavelength bands to be able to determine the magnitude of extinction. Also, High Surface
Brightness regions were avoided in order to minimize the source confusion or crowding. 

\begin{table}
\caption{Numbers of Cepheid Calibrators for Secondary Methods. 
 \label{tbl:F01}} \bigskip 

\begin{tabular}{ccccccccccc}
\hline
Secondary Method & $N(pre-HST)$ & Max.dist & $N (post-HST)$ & Max dist. \\
\hline
Type Ia Supernovae  & 0 & n/a & 06 & 22.4 Mpc \\
Tully-Fisher relation  & 5 & 3.70 Mpc & 21 & 21.5 Mpc \\
Surface brightness flactuation  & 1 & 0.78 Mpc & 06 &19.0 Mpc \\ 
Fundamental Plane  & 0 & n/a & 03 & 22.4 Mpc \\
Type II Supernovae  & 1 & 0.05 Mpc & 04 & 9.75 Mpc \\
\hline\hline
\end{tabular}
\end{table}

Two different softwares (DoPHOT and ALLFRAME) were used by two different group of researchers; 
and they compared their results only at the end of data reduction phase. This double blind approach 
minimizes the systematic uncertainties in this phase.

The final HST Key Project data set consists of 78 data points of five varieties of secondary distance 
indicators(see Table~\ref{tbl:data_set}). Out of which, 36 SNe Ia and 21 Tully-Fisher galaxy clusters and groups are listed in Table 6
and 7 of F01 respectively. 11 galaxy clusters containing Fundamental Plane for 224 early type galaxies 
and six galaxy clusters with SBF measurements are listed in Table 9 and 10 of F01 respectively. Except 
the Fundamental Plane method, the value of Hubble constant obtained from different methods vary slightly. 
Four type II SNe, which are listed in Table 11 of F01 are excluded from our analysis, since SNe II are 
non-standard candles. Instead we have chosen two data points from \cite{sak00}. The complete data set is 
available in \cite[]{mccl07}. In all the cases, recessional velocities have been corrected to the Cosmic 
Microwave Background (CMB) Radiation frame and thus all the $H_0$ values belong to CMB frame. 
F01 find the value of  $H_0$=72$\pm3_r$$\pm7_s$ km/s/Mpc. Table~\ref{tbl:data_set} shows the value 
and  uncertainties(both random and systematics) obtained for each secondary distance indicator (SNe Ia, TF, SBF, FP \& SNe II ).
			
\begin{table}
\begin{center}
\caption{Uncertainties in $H_0$ for Secondary Methods 
 \label{tbl:data_set}} \bigskip 

\begin{tabular}{cccccccc}
\hline
Secondary Method  & No.of data points & Value of $H_0$ & Uncertainties\\
\hline
Type Ia Supernovae  & 36 &71 & $\pm2_r$$\pm6_s$ \\
Tully-Fisher relation  & 21 &71 & $\pm3_r$$\pm7_s$  \\
Surface brightness flactuation  & 06 &70 & $\pm5_r$$\pm6_s$ \\
Fundamental Plane  & 11 & 82 & $\pm6_r$$\pm9_s$\\
Type II Supernovae{$^\dagger$} & 04 & 72 & $\pm9_r$$\pm7_s$\\
\hline\hline
\end{tabular}
\end{center}
{$^\dagger$} \small  Excluded from our analysis , Instead we have chosen two data points for Tully-Fisher relation 
from \cite{sak00}
\end{table}

\section{Methodology: The $\chi$ Statistic and KS Test}
\label{sec:clt}
{\bf Central Limit Theorem:} 
Central Limit Theorem (hereafter CLT) is a fundamental theorem of statistics; and one can hardly overstate 
its importance. To explain the classical CLT \citep{feller}, consider a sequence $\{X_k\}$ with 
$k=1,2,\ldots,n$ of mutually 
independent random variables with a common distribution. Suppose that $\mu$ and $\sigma^2$ are the mean 
and variance of the common distribution; and let $S_n:=\{X_1+X_2+\ldots+X_n\}/n$ be the mean of the sequence. 
According to the classical CLT, $\sqrt{n}(S_n - \mu)$ approximates the normal distribution with zero mean 
and $\sigma^2$ variance, i.e., N(0, $\sigma^2$). CLT is applicable if the random variables have finite mean 
and finite variance. For instance, if the sequence $\{X_k\}$ is drawn from the Cauchy distribution, 
the variance is not finite and hence CLT fails to hold. The classical version of CLT is also known as the 
Lindeberg-Levy CLT; however, some variants are also available. The Lyapunov CLT, for instance, does not 
require the random variables to be identically distributed. 

With the technological advancement the precision 
of the observation has increased enormously. Consequently, the size of the error bars has reduced drastically, 
hence, we expect small errors (finite in size, hence finite variance). A considerably suitable combination of 
above mentioned fact with emergence of sofisticated statistical techniques of data analysis ensures the 
viability of CLT in astronomical observations.

{\bf $\chi$ Statistic:}
Consider the measurement of $H_0$ with true value $H_0^{true}$. The observed value $H_{0i}^{obs}$
in the $i^{th}$ measurement can be expressed as:
\begin{equation}
H_{0i}^{obs} = H_0^{true} \pm \sigma_i ;
\label{eq:funda}
\end{equation}
where $\sigma_i$ stands for error in the $i{th}$ measurement. In the absence of systematic effects 
we expect the average of the errors to be zero, i.e., $\overline{\sigma}_i=0$. One can use appropriate
statistical techniques, such as maximum likelihood, to obtain the best-fit value from the data. In 
this case the best-fit value will be same as the true value, i.e., ${H_0}^{bf} = H_0^{true}$. 
According to CLT, we also expect the errors to follow the Gaussian distribution. If we define:
\begin{equation}
\chi_i = \frac{H_{0i}^{obs} - H_0^{true}}{\sigma_i} ;
\label{eq:chi}
\end{equation}
then, one expects $\chi_i$ to follow the standard normal, i.e., Gaussian
distribution with zero mean and unit standard deviation. However, there could be systematic errors 
involved in the measurement, which would shift the best-fit value away from the true value; and 
thus $\chi_i$ defined in Eq.~\ref{eq:chi} would be biased. If systematic error in the measurement is 
$\epsilon$, Eq.~\ref{eq:funda} is modified to 
\begin{equation}
H_{0i}^{obs} = H_0^{true} \pm \sigma_i + \epsilon \,  .
\label{eq:sys}
\end{equation}
So, the true value in Eq.~\ref{eq:chi} should be replaced with the best-fit value, ${H}_0^{bf}$. 
The equation 
takes the form
\begin{equation}
\chi_i = \frac{H_{0i}^{obs} - {H}_0^{bf}}{\sigma_i} \, ;
\label{eq:chinew}
\end{equation}
If all the measurements in 
the data are statistically uncorrelated then the random variable, $\chi_i$, defined in Eq.~\ref{eq:chinew} 
should follow a standard normal distribution. The method can be easily generalized; one can define $\chi_i$ 
for any physical observable $Y$. If the observed value in the $i^{th}$ measurement is $Y_i$, with uncertainty 
$\sigma_i$, then:
\begin{equation}
\chi_i = \frac{Y_{i} - {Y}^{bf}}{\sigma_i} \, ;
\label{eq:chigen}
\end{equation}
where $Y^{bf}$ is the best-fit value of $Y$. 

{\bf The KS Test:}
Kolmogorov-Smirnov test is a standard tool to determine whether or not a given sample follows the 
Gaussian distribution \citep{numrec}. It compares the cumulative distribution function
\begin{equation}
F(x) = \int_{-\infty}^x f(x)dx
\end{equation}
with the corresponding experimental quantity
\begin{equation}
S(x) = \frac{Number \, of \, observations \, with \, x_i < x}{Total \, Number}
\end{equation}

The test statistic is the maximum difference $k$ between the two functions:
\begin{equation}
k = sup \{F(x) -S(x) \}
\end{equation}

\begin{table}
\begin{center}
\caption{Best-fit value for $H_0$. 
 \label{tbl:h0_bf}} \bigskip 

\begin{tabular}{cccc}
\hline
Best-fit  & $\chi^2$ & $\chi^2_{\rm per \, dof} $ \\
\hline
72.0  & 194.1 & 2.6 \\
\hline\hline
\end{tabular}
\end{center}
\end{table}

\begin{figure}
\centering
\includegraphics[width=0.4\textwidth]{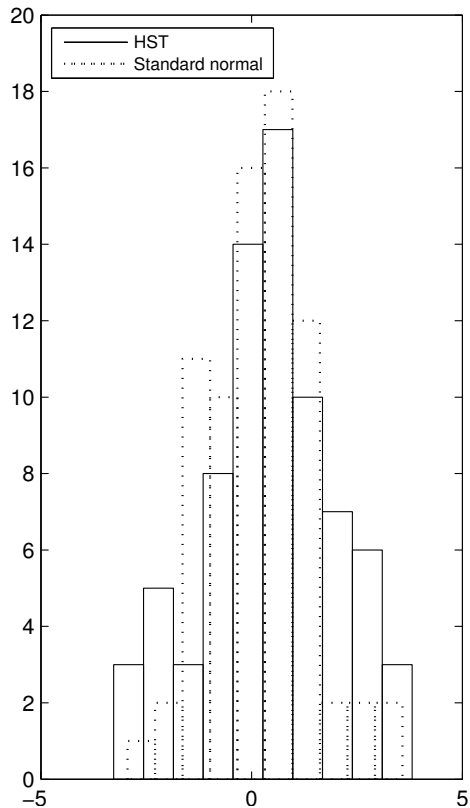}
\caption{Histogram of $\chi_i$`s is compared with that of standard normal distribution.}
\label{fig:hist}
\end{figure}

We set our null hypothesis as: ``The errors in the HST key project data are Gaussian and hence 
$\chi_i$'s in Eq.~\ref{eq:chinew} follow standard normal distribution". We apply KS test to 
calculate the test statistic and the p-value (the probability of obtaining the observed 
sample when the null hypothesis is actually true).

For this, we use Matlab function \emph{kstest[h,p,k,cv]}; where $`k'$ is the maximum distance 
between the two distributions, and $cv$ is the critical value which is decided by the 
significance level ($\alpha$). Different values of $\alpha$, indicate different tolerance 
levels for false rejection of the null hypothesis. For instance, $\alpha=0.01$ means that we 
allow 1\% of the times to reject the null hypothesis when it is actually true. $cv$ is the critical
probability to obtain/generate the data set in question given the null hypothesis.
A value $h=1$ is returned by the test if $p < cv$ and the null hypothesis is rejected.
While for $p>cv$, $h$ remains $0$ and the null hypothesis is not rejected. 

\begin{figure}
\centering
\includegraphics[width=0.60\textwidth]{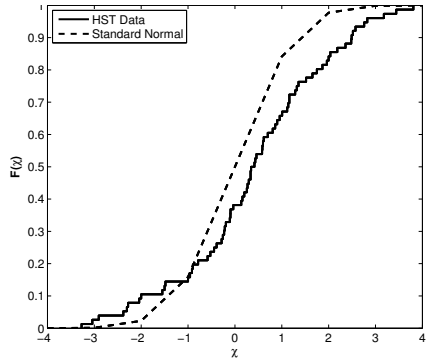}
\caption{A comparison of cumulative distribution of $\chi_i$`s with that of standard normal distribution.}
\label{fig:hstks}
\end{figure}

\section{Results}
We first calculate the best-fit value of The Hubble constant, $H_0^{bf}$ by minimizing $\chi^2$.We obtain $H_0^{bf} =72$
km/s/Mpc, which is shown in Table~\ref{tbl:h0_bf}. The value of $\chi^2$ is too large which suggests that the errors have 
been underestimated. 

As a first check, we calculate $\chi_i$ for each data point and plot a histogram of the values 
in Fig~\ref{fig:hist}. Mean and standard deviation of the $\chi_i$'s are $0.40$ and $1.55$ 
respectively. A histogram of 76 random numbers, generated using the Matlab function "randn", 
is also plotted in the same figure. It is clear from Fig~\ref{fig:hist} that the $\chi_i$`s are 
spread more compared to the standard normal distribution and have thick tails.

\begin{table}
\begin{center}
\caption{Results of KS-test.
 \label{tbl:ksresults}} \bigskip 

\begin{tabular}{cccc}
\hline
$\alpha$ & $cv$ & $p$-value & $ k $  \\
\hline
0.01 & 0.1841 & 0.0048 & 0.1966  \\
0.05 & 0.1534 & 0.0048 & 0.1966  \\
0.10 & 0.1381 & 0.0048 & 0.1966  \\
\hline\hline
\end{tabular}
\end{center}
\end{table}

Results of KS test for $\chi_i$`s are shown in Table~\ref{tbl:ksresults}. The $p$-value is only 0.48\%, 
and is always smaller than $cv$. Thus the null hypothesis is always rejected. Fig.~\ref{fig:hstks} 
shows the cumulative distribution of errors against that of Gaussian distribution. Difference between 
the two distributions is quite visible. Maximum vertical distance is $k=0.1966$. 

\section{Conclusion}
We have presented a neat and simple method to detect the non-Gaussian errors in experimental data; 
and applied it on the HST Key Project data. Our analysis suggests the presence of non-Gaussian 
errors in the HST Key data. The possibility that the non-Gaussian part could be random with some 
other distribution seems unlikely in the light of CLT. The other possibility, that systematic 
effects are making the errors non-Gaussian seems plausible. The systematics could be attributed to 
any one or a combination of the  
following reasons: a) the unknown systematics of the secondary methods; b) zero-point of Cepheid 
P-L relation is not well determined; c) metallicity dependence of the zero-point of P-L relation; 
d) systematic effects arising in the data reduction techniques (in some cases, DoPHOT and ALLFRAME 
give different results); e) calibration of various instruments e.g., complicacy in charge transfer 
efficiency of WFPC2 etc. The detailed treatment of systematics and the method used, could find its
profound impact on improving instrumentation of ongoing and future missions. 

\acknowledgments
MS thanks DMRC for support, AP thanks Amit Sharma for help in typesetting. 
SG thanks Tarun Deep Saini for discussion and colleagues of ASAS for support. 


\end{document}